# Hall effect spintronics for gas detection.

A. Gerber, G. Kopnov and M. Karpovski

Raymond and Beverly Sackler Faculty of Exact Sciences,
School of Physics and Astronomy
Tel Aviv University
Ramat Aviv 69978 Tel Aviv, Israel

We present the concept of magnetic gas detection by the Extraordinary Hall effect (EHE). The technique is compatible with the existing conductometric gas detection technologies and allows simultaneous measurement of two independent parameters: resistivity and magnetization affected by the target gas. Feasibility of the approach is demonstrated by detecting low concentration hydrogen using thin CoPd films as the sensor material. The Hall effect sensitivity of the optimized samples exceeds 240% per $10^4$ ppm at hydrogen concentrations below 0.5% in the hydrogen/nitrogen atmosphere, which is more than two orders of magnitude higher than the sensitivity of the conductance detection.

Reliable detection of hazardous, harmful, or toxic gases has become a major issue due to more stringent environmental and safety regulations worldwide. Solid state conductometric gas sensors present a high potential for applications where the use of conventional analytical systems such as gas chromatography or optical detection is prohibitively expensive or impossible. Operation of these sensors is based on a change of electric conductivity when exposed to an atmosphere containing specific reagents, usually caused by charge transfer between the sensor material and the adsorbed species. Two significant disadvantages of such sensors are the lack of chemical selectivity and sensitivity to humidity. The materials are normally sensitive to more than one chemical species and show cross-sensitivity when different reactive gases are present simultaneously in the atmosphere. When the only parameter measured by the sensor is the change of resistance one can only record the overall electrical effect of quite complex surface reactions. In other words, by only measuring the resistance change one does not have the needed discrimination for the correlation between specific surface species and their electrical effect. In principle, the needed discrimination can be provided by the results obtained by applying additional spectroscopic techniques. Unfortunately, most of the standard spectroscopic investigations are performed in conditions far away from the ones normally encountered in real applications: in ultra-high vacuum, at low temperatures, exposed to high concentrations of reactive gases, etc [1, 2].

The concept of magnetic gas detection has been promoted by a number of researchers that discovered that magnetic properties of several materials are modified when exposed to certain gases, for example to hydrogen. Interaction of hydrogen with ferromagnetic structures containing Pd was shown to change their structural, electronic, optical, and magnetic properties [3]. Significant modifications in susceptibility, magnetization, magnetic anisotropy and ferromagnetic resonance were found in Co/Pd multilayers [4, 5], Pd/Co/Pd tri-layers [6,7], Pd/Fe, Pd/Co and Pd/Ni bilayers [8, 9] and in Pd-rich CoPd alloy films [10, 11]. Additional materials, like Fe/Nb [12] and Fe/V superlattices [13, 14] and $SnFeO_2$ ferrites [15] also demonstrated systematic changes in magnetic properties and exchange coupling when loaded with hydrogen. All effects mentioned above were detected by using laboratory magnetometric techniques and equipment like polarized neutron reflectivity, X-ray resonant magnetic scattering (XRMS), superconducting quantum interference devices (SQUID), vibrating magnetometers, optical Kerr effect and ferromagnetic resonance setups. Adaptation of these techniques to field conditions present a formidable challenge. Therefore, although the idea of gas sensing using the magnetic properties of ferromagnetic materials has been formulated, its realization in practical devices was not implemented so far.

In this Letter, we present the concept of magnetic gas detection using the extraordinary Hall effect [16, 17]. The essence of the effect is the following: electric current flowing along magnetic film generates voltage in direction perpendicular to the current direction, given by:

$$V_H = R_H I = \frac{I}{t}(R_{OHE} B + R_{EHE} \mu_0 M) \tag{1}$$

where $R_H$ is the Hall resistance, $I$ - current, $t$ - thickness of the film, $B$, and $M$ are components of the magnetic induction and magnetization normal to the film plane. $R_{OHE}$ is the ordinary Hall effect coefficient related to the Lorentz force acting on moving charge carriers. $R_{EHE}$, the extraordinary Hall effect coefficient, is associated with a break of the right-left symmetry at spin-orbit scattering in magnetic materials. The EHE contribution can exceed significantly the ordinary Hall effect term in the relevant low field range, and the total Hall resistance $R_H$ can be approximated as:

$$R_H = V_H/I = \mu_0 R_{EHE} M / t \tag{2}$$

Thus, the Hall signal is directly proportional to magnetization. The effect is used as a highly sensitive tool in studies of magnetic properties of ultra-thin magnetic films and nano-structures [18]. Prospects of the Hall effect-based spintronics for magnetic sensors, memories and logic devices [19, 20] were boosted recently by discovery of a huge EHE in amorphous CoFeB oxides, that exhibit the magnetic field sensitivity three orders of magnitude higher than the best achieved in semiconducting materials [21]. Here, we suggest to use the extraordinary Hall effect as a tool for monitoring changes in magnetic properties of the sensor material exposed to specific gaseous elements. The measurement procedure is technically similar to the four-probe measurement of resistance with two modifications: 1) Hall voltage is measured in direction perpendicular to electric current flow, and 2) the measurement is generally done under a bias magnetic field. In the future sensor we envisage, two independent properties:

resistance and EHE will be measured simultaneously in the same magnetotransport setup.

To estimate feasibility of the EHE gas detection we studied the EHE response to hydrogen using thin CoPd alloy films. Hydrogen is highly soluble in palladium, making palladium the metal of choice in hydrogen sensors. The palladium lattice expands significantly with absorption of hydrogen (0.15% in the α- phase and 3.4% in the β-phase), and resistivity of Pd increases with conversion into palladium hydride [22]. Similar response is also observed in Pd-based alloys [23]. Our earlier studies of Co-Pd alloys and multilayers revealed a strong sensitivity of the magnitude and polarity of the EHE signal on the relative content of the system, in particular for Co volume concentrations in the range 10% - 30% [24]. We started this study by assuming that absorption of hydrogen by palladium will modify the structure and electronic state of the system and thus affect the EHE signal.

Polycrystalline $Co_xPd_{1-x}$ films with Co atomic concentration x in the range $0 \leq x \leq 0.4$ were deposited by e-beam co-evaporation from two separate targets on room temperature GaAs substrates. Co and Pd are completely soluble and form an equilibrium fcc solid solution phase at all compositions [25]. Film thickness varied between 5 nm and 20 nm. No post-deposition alloying was used. Several samples were deposited on silicon and glass substrates and demonstrated the response similar to those deposited on GaAs.

Fig.1 presents the Hall resistivity $\rho_H = R_H t$ as a function of magnetic field for four $Co_xPd_{1-x}$ samples with x = 0.08, 0.15, 0.2 and 0.25 (atomic concentration) and thickness 15nm, 18nm, 15nm and 14nm respectively measured in ambient air at room temperature. The ordinary Hall effect, corresponding to the high field linear slope beyond magnetization saturation, is negligible, and the observed signal is mainly due to the EHE. Polarity of the effect, defined as $d\rho_H/dB$, indicates the dominance of the right-hand versus left-hand spin-orbital scattering. The polarity reverses between x = 0.15 and x = 0.2. Samples richer in Co exhibit a positive polarity, while samples richer in Pd have a negative one. The out-of-plane magnetic anisotropy with a significant hysteresis is developed in samples in a vicinity of the EHE sign reversal point. Development of the perpendicular anisotropy has been attributed to a strained state of thin CoPd films [26], that are known to have a very large magnetostriction reaching its maximum in the same concentration range [27, 28]. Both the EHE and magnetostriction are results of the spin-orbit interactions in ferromagnetic materials, however the correlation between the two phenomena is not explained.

Replacement of the ambient air by the pure nitrogen or by the pure carbon monoxide CO atmospheres does not affect the EHE loops. However, the response is significant when hydrogen is added. The four samples shown in Fig.1 were measured in hydrogen/nitrogen $H_2/N_2$ mixture with 4% of hydrogen, and the results are presented in Fig.2. The magnitude of the saturated EHE signal is reduced by 2% to 15% in all hydrogenated samples. The most pronounced changes are observed in the hysteresis loops of the samples with the out-of-plane anisotropy. Width of the quadratic hysteresis loop shrinks in the $Co_{0.15}Pd_{0.85}$ sample (Fig.2b). The $Co_{0.2}Pd_{0.8}$ sample also demonstrates a reduction of the coercive field together with the zero field remanence signal reduced to about a half (Fig.2c). Reduction of the coercive field and the remanence indicate the decreasing perpendicular magnetic anisotropy with hydrogen absorption.

Fig.3 presents the field dependent hysteresis loops measured in 5 nm thick $Co_{0.17}Pd_{0.83}$ sample in $H_2/N_2$ atmosphere at different hydrogen concentrations between 0 and 4%. Thinner films seem to be attractive for sensing purposes due to a higher surface to volume ratio, and since the absolute value of the measured signal (Eqs.1 and 2) increases both by the reducing thickness $t$ and by enhancing the EHE coefficient $R_{EHE}$ boosted by the spin-orbit surface scattering [29]. After the initial measurement in $N_2$ (99.998%) at atmospheric pressure, the sample chamber was filled with $H_2$ 4% $H_2/N_2$ mixture. The following sequence of measurements at reduced hydrogen concentrations was done after pumping the chamber to half of atmospheric pressure and refilling the chamber by nitrogen. After completing the sequence, the sample was re-measured in $N_2$. The hysteresis loops are fully reproducible when the sequence is repeated. As seen, the saturated magnitude of the signal at high field, the remanence at zero field and the width of the hysteresis loop decrease with increasing hydrogen concentration.

The quantitative data are shown in Fig.4. Fig.4a presents the coercive field $H_c$ as a function of hydrogen concentration $y$. $H_c$ can be well presented by the power law dependence on the hydrogen concentration as: $H_c(y) = H_c(0)y^{-\gamma}$ with $H_c(0) \approx 7\ mT$ and $\gamma \approx 0.3$, i.e. it varies significantly at low hydrogen concentrations and saturates towards 4%. Fig. 4b presents the normalized change of the EHE signal measured at several fixed fields within the hysteresis loop (H = 0, 1.5 mT and 4 mT). The normalized EHE change is defined as: $\Delta R_{H,norm}(y) = \frac{\Delta R_H(y)}{R_H(0)} = \frac{R_H(y) - R_H(0)}{R_H(0)}$, where $y$ is the hydrogen concentration. The signal varies strongly at low $H_2$ concentrations and saturates by approaching 4%. The rate of the signal variation and the range of the linear response depend on the bias field. At 4 mT bias field the sensitivity ($S = d\Delta R_{H,norm}/dy$) exceeds 240%/$10^4$ $H_2$ ppm at hydrogen concentrations below 0.5%. At 1.5 mT the sensitive range extends up to 2% of hydrogen with sensitivity about 30%/$10^4$ ppm. Variation of the remnant EHE signal at zero bias field reaches 30% at 4% hydrogen. The response is not linear over a wider concentration range, which should be taken into account in calibration of the future sensors.

Resistance response to hydrogen measured simultaneously with the EHE is shown in Fig.4c. The data taken at zero field (○) and in the magnetically saturated state under 0.1T bias field (×) are presented in the form of the normalized resistance change, defined as: $\Delta R_{norm} = \frac{R(y) - R(0)}{R(0)}$. Both at zero and under 0.1T field resistance increases about linearly with hydrogen concentration up to 4%. The resistance sensitivity to hydrogen concentration, defined as: $d\Delta R_{norm}/dy$, is about 0.8%/$10^4$ ppm. Magnetoresistance of the sample is small, negative and independent on hydrogen absorption. Therefore, the resistance changes caused by hydrogen adsorption don't depend on the bias field. Fig.4c also presents the normalized EHE response $\Delta R_{H,norm}$ measured in the magnetically saturated state at a fixed field 0.1T. The EHE and the resistivity responses to hydrogen absorption are independent of each other. The magnetic EHE response is negative, reaches 12% at low hydrogen presence and saturates towards 4% concentration. Resistivity increases in the measured $H_2$ concentration range with no signs of saturation. Following Eq.2, the EHE signal depends on the EHE coefficient $R_{EHE}$ and magnetization M. $R_{EHE}$ scales with resistivity as: $R_{EHE} \propto \rho$ due to the skew scattering mechanism or as $R_{EHE} \propto \rho^2$, following the intrinsic Berry phase mechanism or the extrinsic mechanism of side jump scattering [30]. Changes of the saturated magnetization and of the field dependent hysteresis loop due to gas absorption are uncorrelated with resistivity, which makes the EHE and resistivity responses independent.

Reduction of the saturated EHE signal with increasing hydrogen concentration is consistent with the generally observed decrease of the total magnetization [5, 10] in hydrogenated Co/Pd systems, the effect attributed to modification of the electronic structure of the material. On the other hand, the effect of hydrogen absorption on the perpendicular anisotropy is ambivalent. Enhancement of the perpendicular magnetic anisotropy was found in hydrogenated Pd/Co/Pd trilayers [6], associated by the authors with improvements of Pd (1,1,1) orientation, and in Pd-rich alloy film [10], attributed to the development of a long-range magnetic order. The coercive field and the perpendicular magnetic anisotropy of our samples decrease with hydrogen absorption, similar to Co/Pd multilayers reported in Ref. [5]. We tentatively suggest that changes in magnetic anisotropy depend strongly on magnetostriction and strain of the material, similar to the concentration dependence of non-hydrogenated CoPd films (Fig.1). More studies are needed to clarify this point.

To summarize, one can expect that selectivity of solid state gas sensors will be improved by extending the range of independent measurable parameters complementing the conductometric sensing. We argue that the extraordinary Hall effect (EHE), sensitive to variations of magnetic properties of ferromagnetic materials, can serve as such complementary magnetotransport parameter. Possibility to apply the technique for gas sensing was demonstrated by detecting low concentration hydrogen using thin CoPd films. Sensitivity of the EHE response in the optimized samples exceeds 240% per $10^4$ ppm at hydrogen concentrations below 0.5% in the hydrogen/nitrogen atmosphere, which is more than two orders of magnitude higher than the sensitivity of the conductance detection.

The research was supported by the State of Israel Ministry of Science, Technology and Space grant No.53453.

**Figure captions.**

Fig.1. Field dependence of the Hall resistivity of $Co_xPd_{1-x}$ films with x = 0.08, 0.15, 0.2 and 0.25 (atomic concentrations) and respective thickness 15 nm, 18 nm, 15 nm and 14 nm measured in ambient air at room temperature.

Fig.2. Hall effect resistance as a function of magnetic field of $Co_xPd_{1-x}$ films measured in air (open circles) and in hydrogen/nitrogen $H_2/N_2$ mixture with 4% of hydrogen (solid circles): (a) $Co_{0.08}Pd_{0.92}$; (b) $Co_{0.15}Pd_{0.85}$; (c) $Co_{0.2}Pd_{0.8}$; (d) $Co_{0.25}Pd_{0.75}$.

Fig.3. EHE resistance hysteresis loops measured in 5 nm thick $Co_{0.17}Pd_{0.83}$ film in $H_2/N_2$ atmosphere with different $H_2$ concentrations (y = 0%, 0.125%, 0.25%, 0.5%, 1%, 2% and 4%).

Fig.4. Hydrogen concentration dependence of: (a) the coercive field; (b) the normalized EHE change $\Delta R_{H,norm}$ under bias fields 0 mT, 1.5 mT and 4 mT within the hysteresis loop; (c) the normalized resistance change $\Delta R_{norm}$ at zero field (○) and under 0.1T bias field (×) – right vertical axis, and the normalized EHE change $\Delta R_{H,norm}$ under 0.1T bias field – left vertical axis, measured in 5 nm thick $Co_{0.17}Pd_{0.83}$ film.

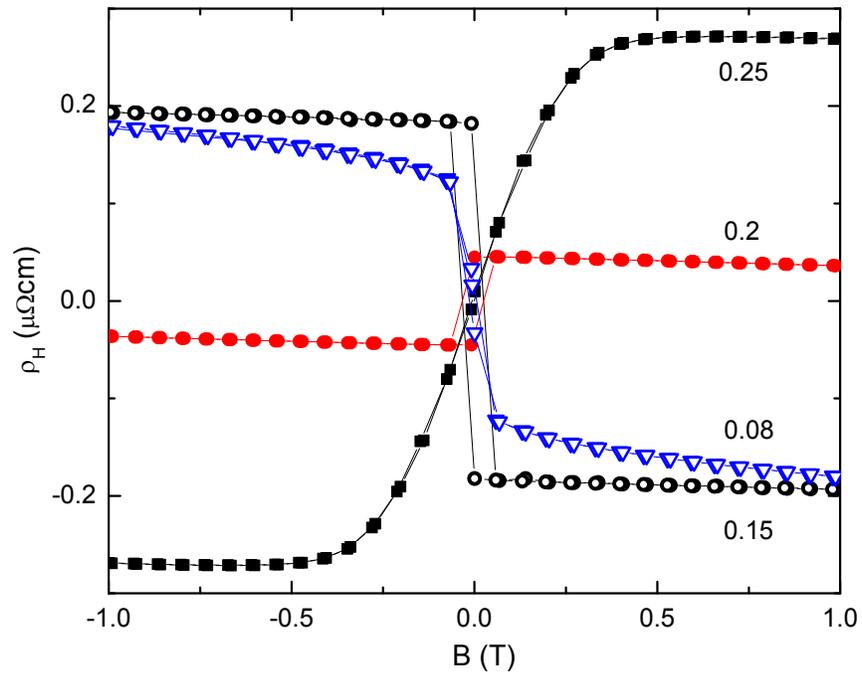

Fig.1

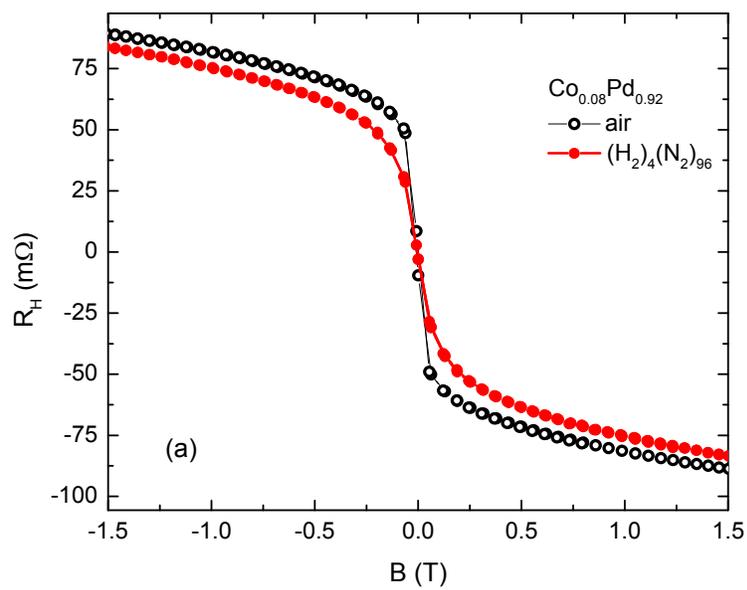

Fig. 2a

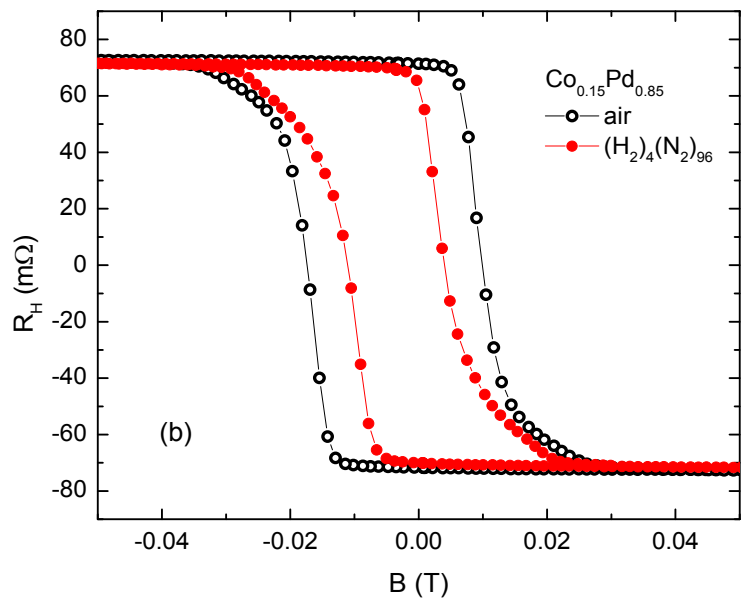

Fig. 2b

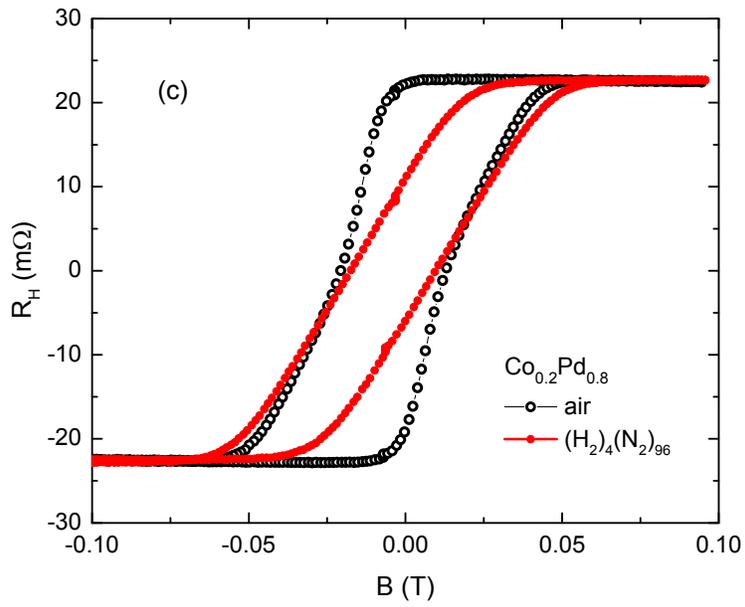

Fig. 2c

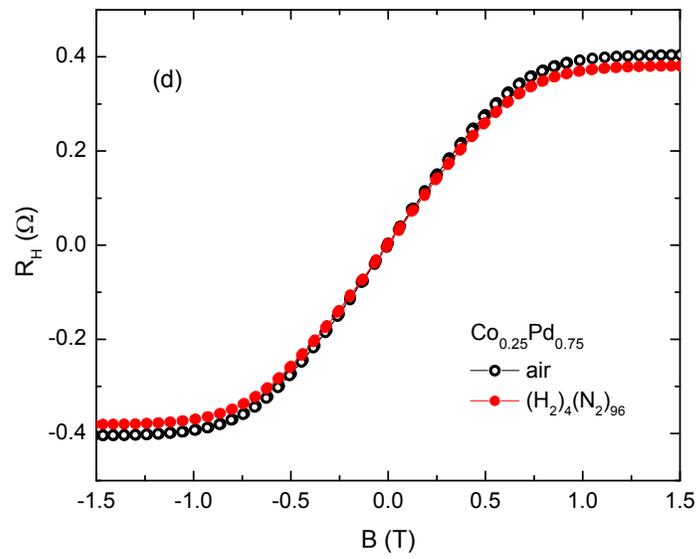

Fig. 2d

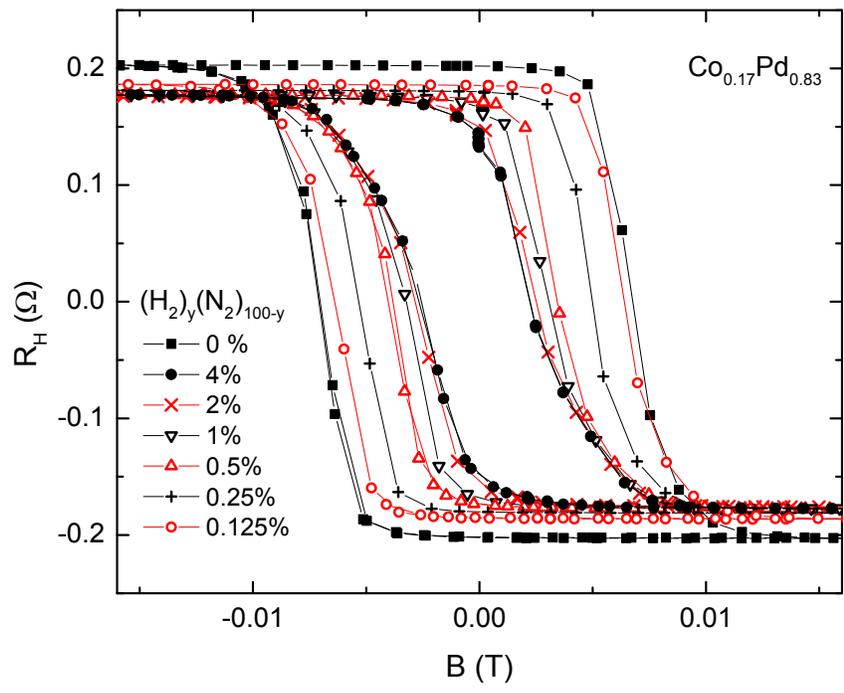

Fig. 3

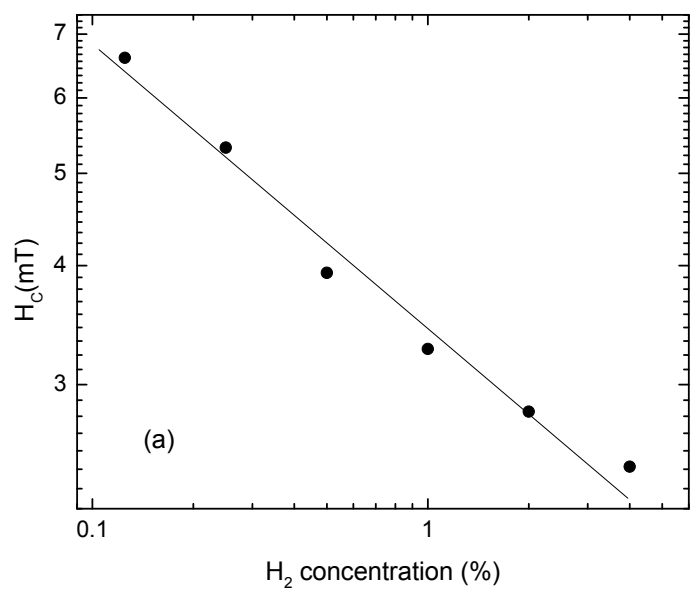

Fig. 4a

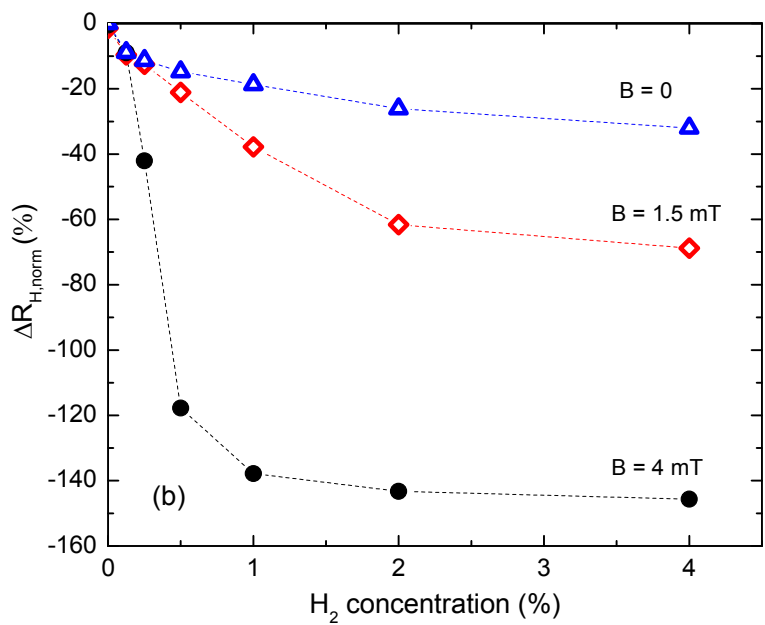

Fig. 4b

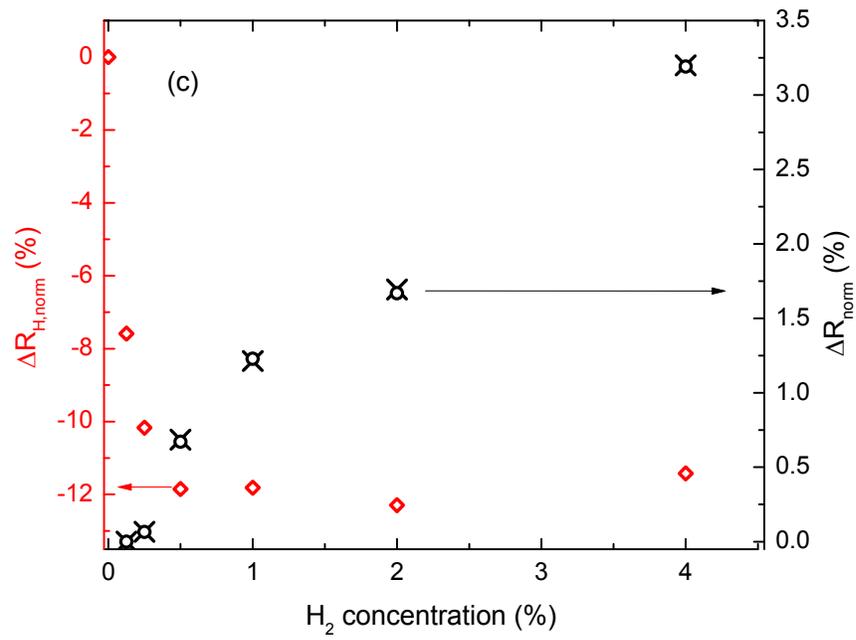

Fig. 4c

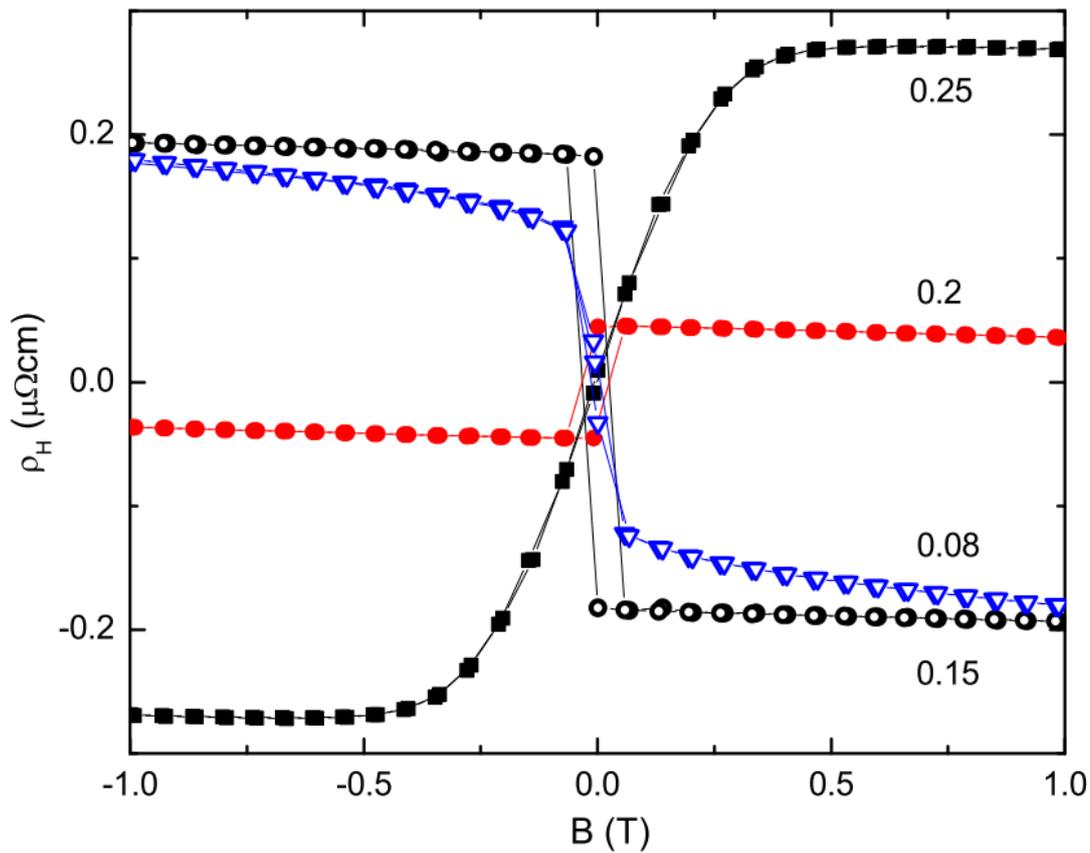

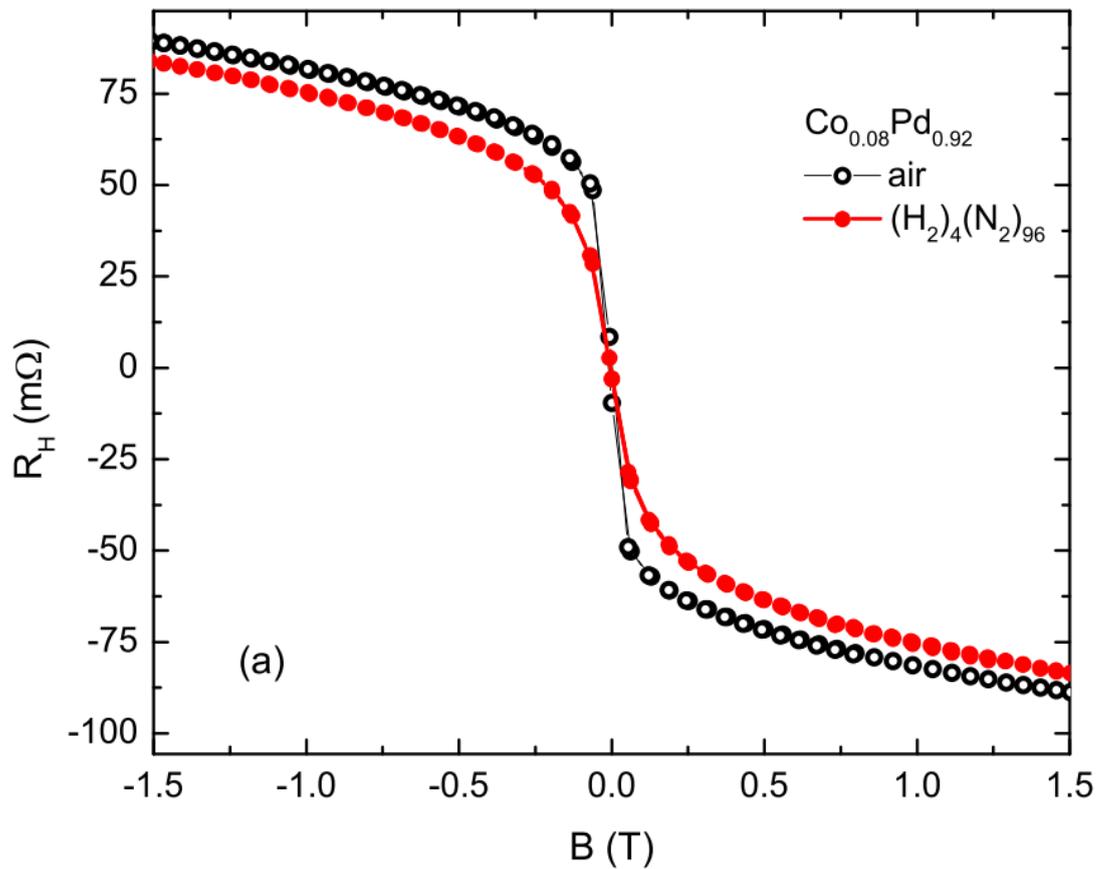

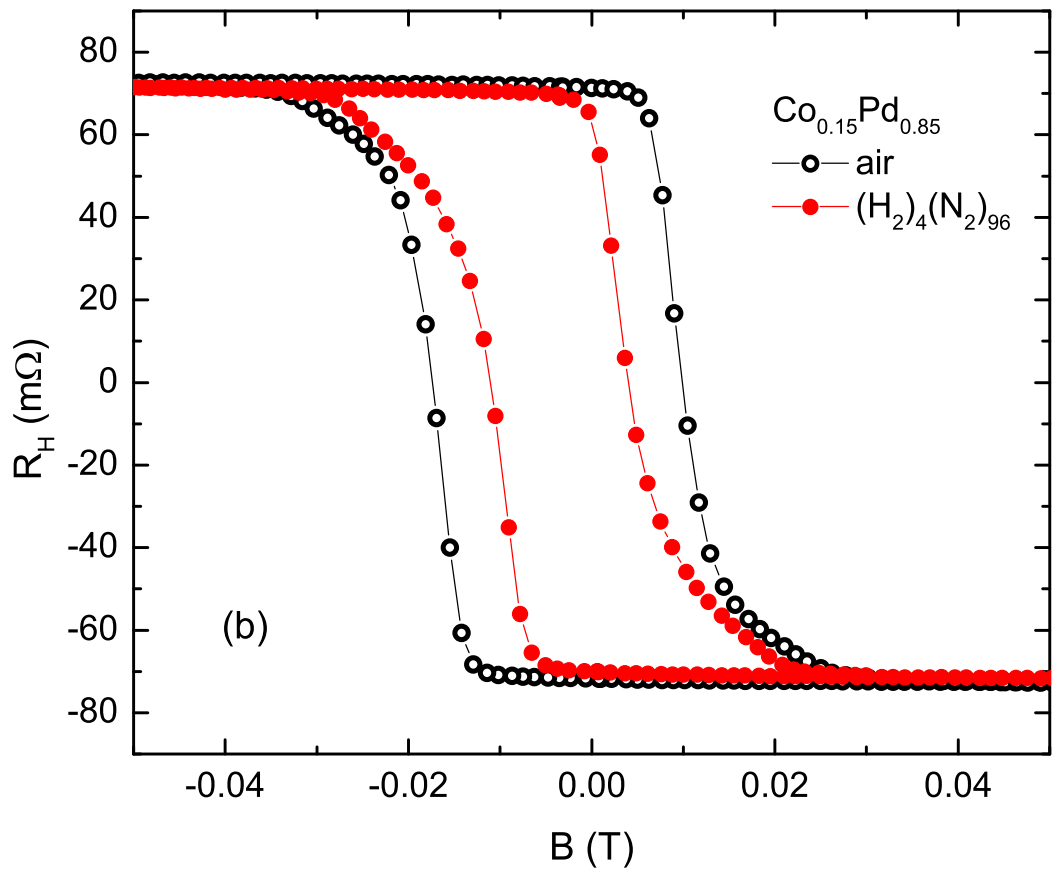

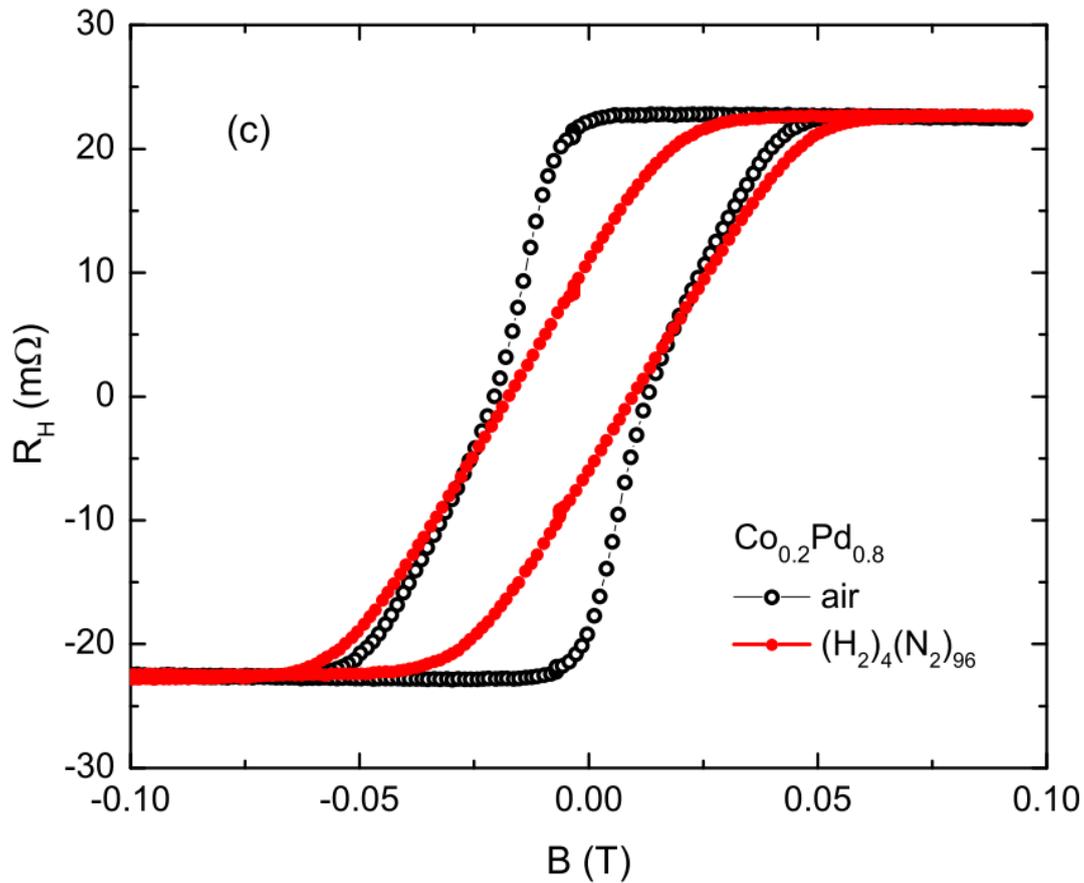

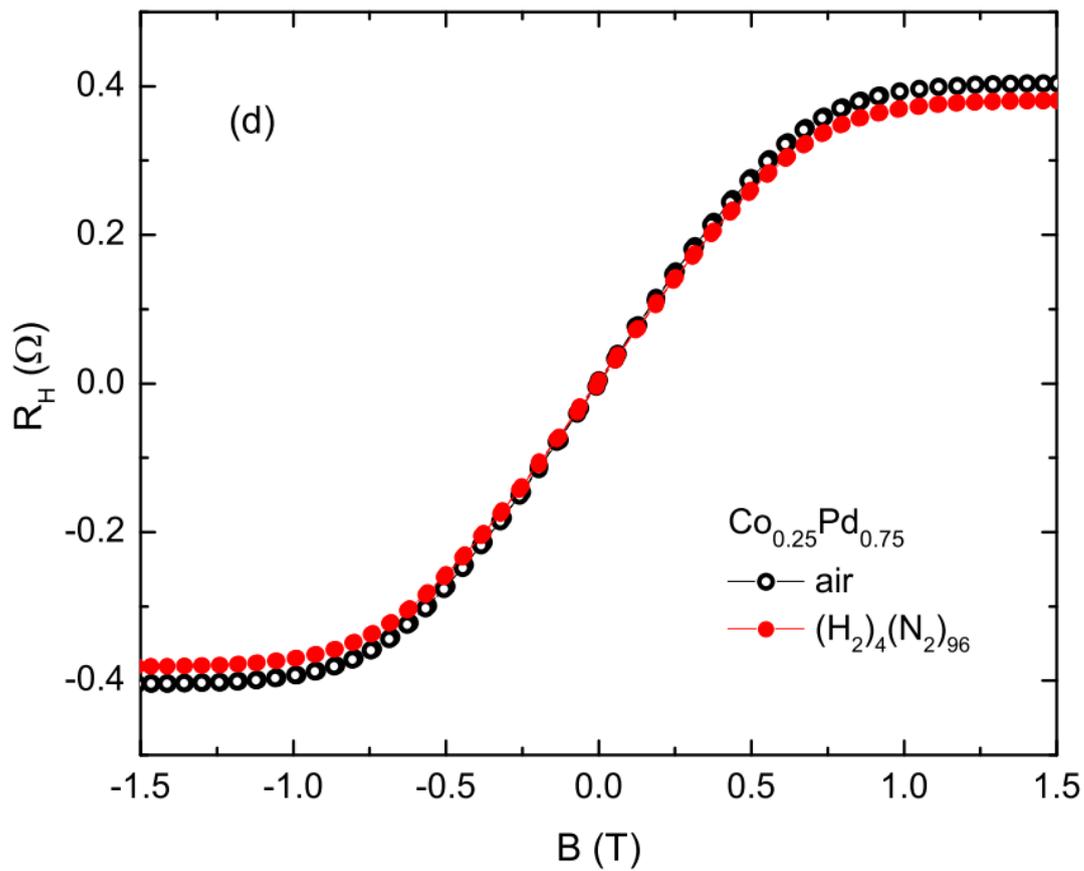

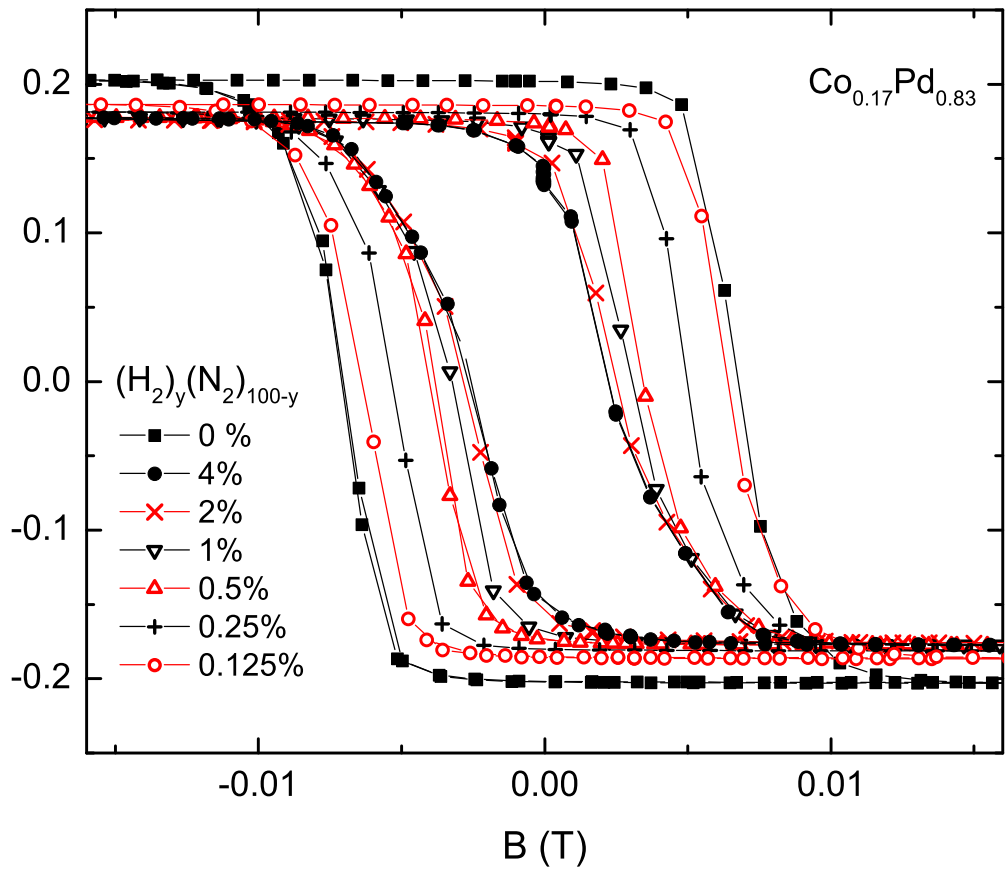

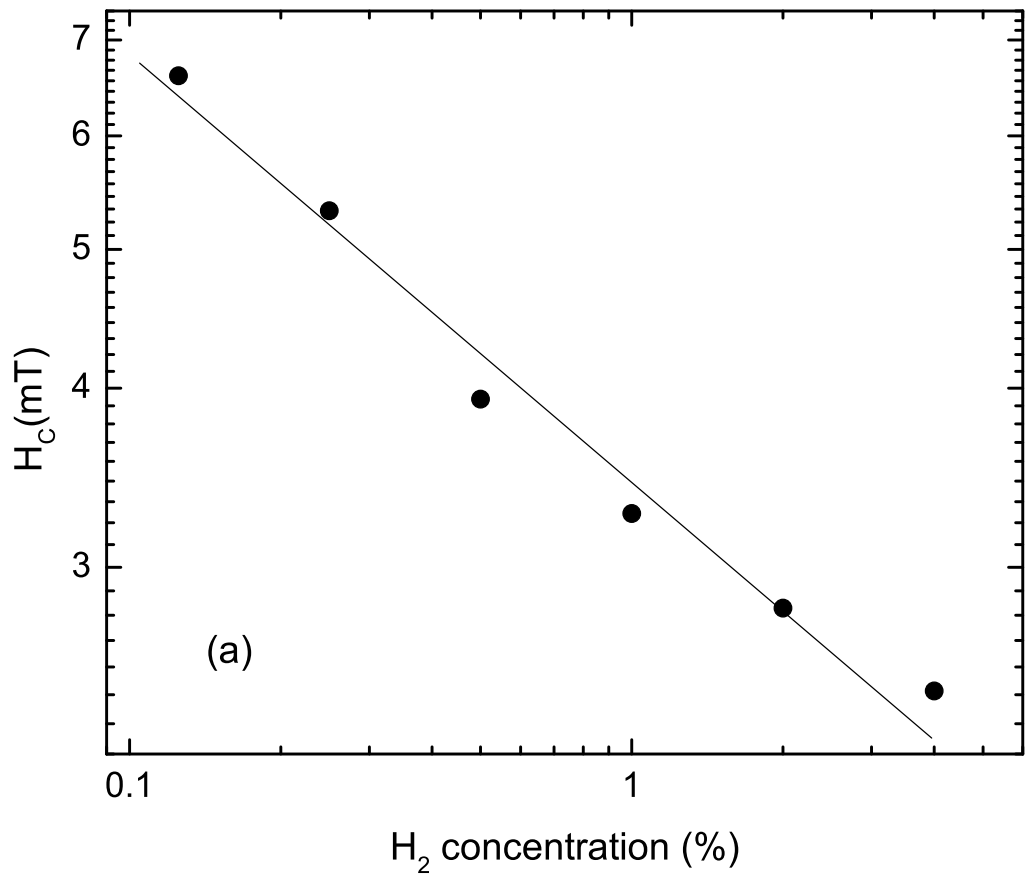

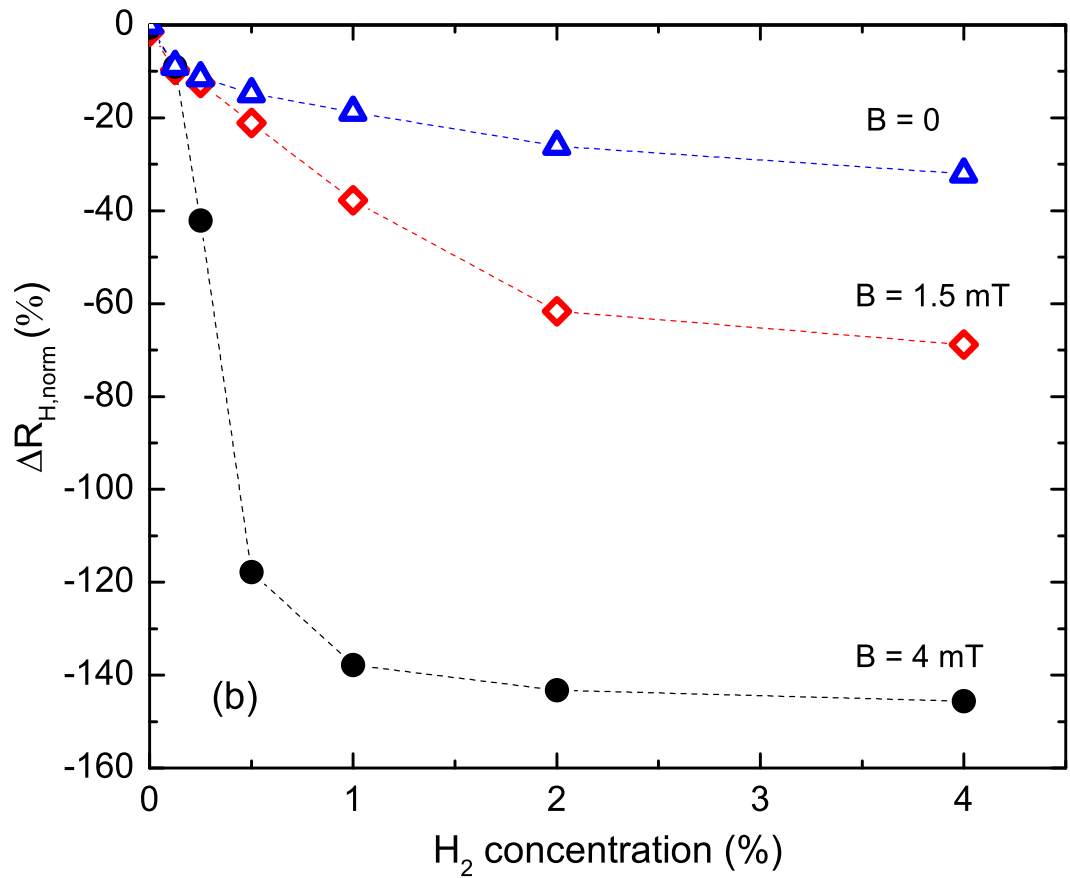

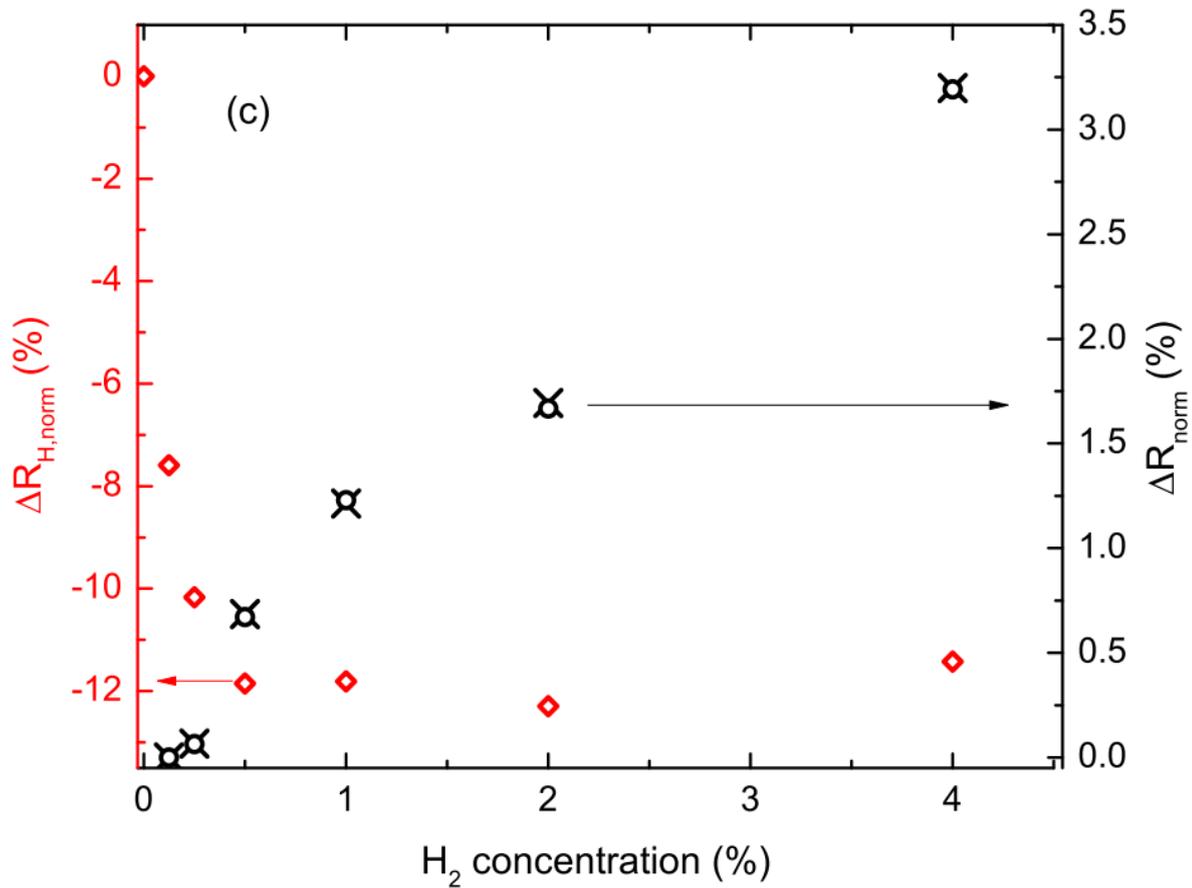